# HAS THE SECOND LAW OF THERMODYNAMICS REALLY BEEN VIOLATED?

Leonardo Chiatti
ASL VT Medical Physics Laboratory
Via San Lorenzo 101
01100 Viterbo
Italy

fisica1.san@asl.vt.it

**Abstract**

Some works have appeared in recent accredited literature considering the possibility of macroscopic violations of the second law in simulated as well as really executed experiments. We argue the inexistence of such violations in experiments based on the so-called nonbias diode, demonstrating that the interpretation of the authors is flawed by a confusion between the concepts of thermal equilibrium and thermodynamic equilibrium. We also discuss an isomorphic experimental set up based on the evaporation-condensation of a liquid in a closed atmosphere.
Some critical observations are then made concerning recent attempts to reformulate classical equilibrium thermodynamics.

*Keywords: thermodynamic equilibrium, violations of the second law, nonbias diode*

*PACS : 05.70.-q*

## 1. Introduction

Recently at least two really performed experiments have been published that, in the opinion of the authors, would indicate a violation of the second law on a macroscopic scale. These presumed violations did not occur at the quantum or cosmological limit, but under ordinary conditions of classical phenomenology observable in the laboratory; thus the validity of classical equilibrium thermodynamics in its current formulation is questioned.
Yelin (1) created a set of 694 identical vacuum diodes connected in parallel, thermally coupled with the external environment but optically and electromagnetically isolated from it. The set of diodes is not powered by external voltages or currents, nevertheless a current on the order of $10^{-8}$ A flows in the short-circuited apparatus due to the effect of rectification of the thermal noise current fluctuating in the apparatus at ambient temperature. The short-circuit current varies with the ambient temperature and, when the external environment is replaced by a thermostat, the relation between current and temperature proves to be definite and reproducible (Fig. 4-4 of ref. 1).
The single "nonbias diode" is realized with particular expedients that allow to obtain a good thermoionic emission at ambient temperature. The ratio of cathodic and anodic surfaces and that of extraction works are chosen so as to ensure, at ambient temperature, a flow of thermoelectrons from the cathode to the anode greater than the flow in the opposite direction from the anode to the cathode. Consequently, at ambient temperature a net electric current flows from the cathode to the anode when the circuit is closed. The control of the optical and electromagnetic isolation of the apparatus is obviously essential in order to eliminate induced currents generated by spurious external fields.

If, instead of short-circuiting the ends, these are closed on a resistive load, there are two distinct possibilities. If the load consists of a passive resistor, heat can be generated by the Joule effect at a temperature higher than the ambient one; if instead it consists of an active load (for example, an electric motor), then a conversion of electric energy into another form of energy or work (for example, mechanical work) is possible. According to Yelin, both these circumstances would entail a violation of the second law. The Yelin experiment was subsequently updated (2), obtaining currents on the order of μA.

Sheehan (3) provided a survey of experiments that would lead to violations of the second law. The first of these experiments, which has been really performed (3,4,5,6,7), is practically identical to that performed by Yelin; Sheehan does not cite Yelin probably due to the limited diffusion of Yelin's work in English. The diagram of the apparatus shown in Fig. 1 of ref. (3) is practically the same as that of the nonbias diode of Yelin, except for secondary technical characteristics. In the next section we will discuss the analysis of the Yelin-Sheehan experiment.

The other three setups proposed by Sheehan (but not realized) are not particularly interesting, in our opinion, as we are unable to see any potential violation of the second law. In the first setup, a piston moves by the reaction due to differential emission of charges from its walls; therefore the source of the force is the piston itself, not the gas of charge particles surrounding it. Attributing the work performed on the piston to the gas is like attributing to the air the work performed on the jet engine of an airplane, which in reality moves due to the reaction to the expelled jet.

Since the work $L$ is not performed by the gas, the gas does not absorb the heat $Q$ from the thermostat converting it into work $L$, as Sheehan seems to assume; the piston is not a heat engine and thermodynamics is not involved.

The second setup is also based on the reaction of a piston, and the same objections hold for this setup.

The third setup is based on a different type of confusion. If a bar immersed in a gas in conditions of thermal equilibrium absorbs more impulse from the gas on one of its ends than on the other, the bar recoils in the direction of the opposite end. The bar, assumed to be rigid, is therefore subjected to the action of the following external forces: the pressure of the gas on the two ends, its own weight, the viscous friction of the gas. After an initial transient, the system will reach a configuration of equilibrium between all these forces, and the bar will move with uniform motion. Thus, apart from the initial transient which doesn't concern us in this discussion, the work performed by the external forces on the bar is $L = 0$.

On the other hand, it is implicit that the gas does not transfer any heat to the bar; otherwise the bar would heat up, whereas it is assumed that it remains in thermal equilibrium with the gas. Therefore the heat drawn from the thermostat (a part the transient) is $Q = 0$. From the first law we have a variation in the internal energy of the bar equal to $\varDelta U = Q - L = 0 - 0 = 0$. Once again, we do not have a thermodynamic transformation to which to apply the laws of thermodynamics.

## 2. Comments on the Yelin-Sheehan experiment

Ideally, the setup realized by Yelin and Sheehan can be represented with an electronic circuit composed of a single loop, with a vacuum diode and a resistance $R$ in series. This simple apparatus is supposed to be in thermal equilibrium with the external environment at the absolute temperature $T$. Thus the electric current $I$, derived from the rectification of the thermoionic current inside the diode, flows in the circuit. Upon superficial examination the production of this current, which could be used to power a motor, would seem to come at the expense of the environmental heat. If this were so, there would effectively be the integral conversion into work of the heat absorbed from an external source at the uniform temperature $T$, in clear violation of the second law. However, a simple analysis of the two subcases $R = 0$, $R > 0$ clarifies the problem.

1) Case $R = 0$.

The voltage at the ends of the resistor is $V = RI = 0$; the power dissipated in the resistor is therefore $VI = 0$. Consequently, the work $L$ performed by an electron in a complete loop of the circuit is 0, because no work is actually performed. The contribution of this electron to the total variation of internal energy of the system during the path of the entire loop is $\Delta U = (-e)\Delta V = (-e) \times 0 = 0$, since the total voltage variation along the entire loop is, by Kirchoff's laws, 0. By the first law, the contribution of the electron to the heat absorbed from the environment is therefore $Q = \Delta U + L = 0 + 0 = 0$. Thus there is no real absorption of heat from the environment nor its subsequent conversion into work. If, in agreement with the current convention, we call "heat engine" any device able to absorb heat from one or more sources and convert it, at least partially, into work, this apparatus is not a heat engine. Kelvin and Clausius statements, which apply to heat engines, thus cannot be applied to this apparatus, nor disproved by its operation.

2) Case $R > 0$.

In this case the contribution of the electron to the total variation of internal energy of the system during the path of the entire loop is also, by Kirchoff's laws, $\Delta U = (-e)\Delta V = (-e) \times 0 = 0$. Nevertheless, now the current $I$, flowing across the load, induces a finite potential difference $V = RI$ between cathode and anode (Ohm's "inverse" law). This current develops, crossing the load, a finite power $VI$. The electric field performs a work $L = (-e)(-V)$ on the electron as it crosses the diode; conversely, it performs a work $-L = (-e)V$ on the same electron when it crosses the load. If the load is active (an electric motor, for example) then this work is transferred to the load; thus we have the conversion of electric work into mechanical work or another form of work. If instead the load is passive, this work is dissipated in heat inside the load; in fact from $\Delta U = 0$ follows $Q = L = eV$, that is a total conversion of electric work into heat. Once again, therefore, there is no conversion of environmental heat into work and thus the second law is not disputable with this setup.

The mistake is the following: it is true that there exists a thermal interaction between the environment at temperature $T$ and the gas of thermoelectrons of the diode, but in this interaction there is no net exchange of heat between electrons and the environment. Such an exchange could occur only if the electrons returned to the cathode at a lower temperature than when they left, for which they would absorb internal energy from the environment in order to readjust their energy distribution at equilibrium with the ambient temperature (8). But we have seen that $\Delta U = 0$. The thermal interaction with the environment thus has merely the function of maintaining the temperature of the cathode and anode at a constant level (equal to that of the ambient temperature); in other words the environment plays the role of a *thermostat*.

It should also be noted that we have three systems in interaction:

- the environment and the gas of electrons in the case $R = 0$;
- the environment, the gas of electrons and the load in the case $R > 0$.

In the first case, the two systems return to the same condition at the end of a cycle (loop of the circuit by the electron) and no work is performed.

In the second case, the first two systems return to the same condition at the end of the cycle, but not the load. This in fact heats up if it is a resistor, while it modifies its configuration if it is a conservative load (a motor, for example). In this case heat or work can be generated.

The entire question could therefore be dismissed as a gross error in the physical interpretation of experimental data. We however believe that there is at least one aspect worthy of reconsideration in the Yelin-Sheehan experiment. We note, in fact, that in the particular case $R = 0$ no component of the apparatus exchanges heat or work with the environment, which could therefore be replaced by a finite thermal capacity, ideally thermally isolated, sufficiently large to maintain the temperature of the gas of electrons constant. This thermal reservoir would in fact be a component of the apparatus. Thus we reach the interesting conclusion that **in an isolated system in thermal equilibrium, it is**

**possible to have the spontaneous formation of steady flows, in this case the rectified electric current *I*.**
This "spontaneous generation of order" deserves a comment. A system is said to be in thermal equilibrium if, in its current physical state, a quantity 'temperature' is defined and the value of this quantity is homogenous over the entire system. A system is said to be in thermodynamic equilibrium if, in its current physical state, a set of "state variables" is defined, one of which being the temperature, which all take on homogenous values over the entire system. Thus there can be the case of systems that are in thermal equilibrium but not in thermodynamic equilibrium. It is known that static gradients of the state variables may arise in these systems. An elementary example is that of a motionless gas in thermal equilibrium at the temperature *T* in a constant and homogeneous gravitational field; in this gas there is a pressure gradient along the direction of the gravity field, expressed by the well known barometric formula. One can therefore wonder if in the presence of thermal equilibrium but not thermodynamic equilibrium, there may also arise steady flows, such as flows of fluid or electric currents. The Yelin-Sheehan experiment affirmatively answers to this second question. Now, the second law of thermodynamics does not entirely prohibit self-ordering phenomena in isolated systems in thermal equilibrium such as those that lead to the appearance of such steady flows or static gradients, **except in the case of thermodynamic equilibrium**. It seems that this simple observation is not taken suitably into consideration in some recent attempts to reformulate the classical equilibrium thermodynamics, as we will see in the conclusions.

### 3. An analogue of the Yelin-Sheehan experiment

It is possible to formulate an experiment structurally similar to that of Yelin-Sheehan, based on evaporation-condensation phenomena of a liquid in a closed atmosphere (9). Recall that if a liquid is bounded by a curved surface with principal radii of curvature $r_1$ and $r_2$, the pressure *P* of the liquid on the concave side is expressed as :

$$P = P_0 - \tau \left( \frac{1}{r_1} + \frac{1}{r_2} \right) \qquad (1)$$

where $P_0$ is the value that the pressure would have in the case of a flat surface and $\tau$ is the surface tension. In eq. (1) the radius of curvature has been assumed to be external to the liquid, as in the case of a meniscus in a capillary tube (otherwise the sign – should be replaced with the sign +). Now, the pressure of the liquid determines, in the conditions of thermal equilibrium at the absolute temperature *T*, the saturated vapour pressure of the liquid itself, in agreement with the Clapeyron equation :

$$\ln \frac{e}{e_0} = V \frac{(P - P_0)}{RT} \qquad (2)$$

where *e* and $e_0$ are the vapour pressures corresponding respectively to *P* and $P_0$ values of liquid pressure, *V* is the molar volume of the liquid and *R* is the gas constant. Substituting equation (2) into equation (1), we obtain the expression of the saturated vapour pressure in the case of a concave surface :

$$e = e_0 \exp\left[ -\frac{V\tau}{RT} \left( \frac{1}{r_1} + \frac{1}{r_2} \right) \right] \qquad (3)$$

and, as can be seen, the vapour pressure $e$ is always less than its value $e_0$ corresponding to the flat surface. Now let us consider two containers containing the same liquid, separated in a closed container, thermostated, at two different heights, and let be $z$ the height difference between the two liquid surfaces. We assume that the surface of the liquid in the lower container is flat, while the surface of the liquid in the container located in the higher position is concave [this can be realized with a capillary structure obtained, for example, by immersing microfibre strips in the higher container (9)].

With these premises, the vapour pressure induced by the lower liquid at the height of the higher liquid surface is :

$$e_{low}(z) = e_0 \exp\left[-\frac{Mgz}{RT}\right] \qquad (4)$$

where $M$ is the molecular weight of the liquid and $g$ is the acceleration of gravity. If the higher liquid surface were flat, its vapour pressure would be $e_{high}(z) = e_0$, thus greater than $e_{low}(z)$, with the result that the liquid positioned higher would tend to go out of the upper container to condense in the lower container. But since the higher liquid surface is actually concave, it is possible to choose the setup parameters so that it has $e_{high}(z) < e_{low}(z)$, with consequent condensation of the liquid in the container positioned higher.

Thus the following circuit scheme is conceivable : the liquid collected in the upper container, and that would tend to level the meniscus compensating the effect, is instead made to drip into the lower container by gravity. Thus a liquid circulation forms that is *apparently* sustained by the absorption of environmental heat. In reality, the heat absorbed by a liquid mass that evaporates from the lower container is exactly equal and opposite to the heat released by this same mass when it condensates on the upper container. Furthermore, the work performed by gravity on that mass during its raising is exactly equal and opposite to the work performed by gravity on the same mass condensed into drops when these drops fall into the container below. Thus the heat exchanged by that mass with the thermostat during the entire raising-lowering cycle is $Q = 0$, and the work performed on that mass by the gravity field during the same cycle is $L = 0$. Once again, there is no conversion of environmental heat into work. If the falling drops are made to strike the vanes of a mill, the work performed on the mill derives in effect from a partial conversion of gravitational potential energy, and not from environmental heat.

**4. Conclusions**

Both the experiments carried out by Yelin and Sheehan as well as that proposed by Vignati refer to the formation of steady flows (respectively : an electric current and a fluid current) in systems that are in thermal equilibrium at a certain temperature but are not in thermodynamic equilibrium. In both cases the environment can be replaced, as a simple thermostat, by a thermal reservoir with sufficiently high capacity at the same temperature. This reservoir can be considered as a component of the apparatus, and we thus have a formation – possibly spontaneous – of stationary flows in isolated systems. In conclusion, therefore, these experiments simply show not that the second law is violated, but that in isolated systems order can be self-generated in the form of steady flows, in addition to static gradients. These self-ordering, or self-sustained ordering, phenomena do not violate the second law since this law prohibits them only in the more restrictive case of a system in thermodynamic equilibrium.

From these considerations it clearly follows that an isolated system in thermal equilibrium does not necessarily tend toward a state of thermodynamic equilibrium. The inevitability of this "heat death" constitutes a superficial interpretation of the second law which is, however, also at the basis of a recent proposal for the reformulation of classical equilibrium thermodynamics, that of Brown and

Uffink (10). These authors propose to replace the second law with a weaker, time reversal, version. They propose then to add, to the usual thermodynamic laws (zeroth law, first law, the second law in its revised form, and the third law) a new law that they call "the minus first law". This law would contain the aspects more properly connected with the temporal irreversibility of thermodynamic evolution, usually expressed by the second law in its standard form. Without going into the details of the overall proposal, we limit ourselves to quoting the statement of the minus first law, which is the following : *"When an isolated system finds itself in an arbitrary initial state within a finite fixed volume, it will spontaneously attain a unique state of equilibrium"*. Here "state of equilibrium" must be understood as a state of *thermodynamic* equilibrium, as the authors themselves clarify in the subsequent point (A) of their article :

*(A) [We assume] the existence of equilibrium states for isolated systems. The defining property of such states is that once they are attained, the independent thermodynamic coordinates of the system are spatially homogeneous and remain thereafter constant in time, unless the external conditions are changed. The claim that such states exist is not trivial—it rules out the possibility of spontaneous fluctuation phenomena.*

Actually, it is precisely the dominance of fluctuation phenomena in experiments like those of Yelin-Sheehan (noise current) or Vignati (evaporation) that create conditions which render the minus first law too restrictive to be applicable to the real world. This law is violated even in the case of simple static gradients such as the vertical barometric gradient. This confirms the hypothesis that the second law *truly* and *per se* expresses a time asymmetry, contrary to the opinion of the aforesaid authors and in agreement with the more common views.

**Acknowledgements**

The author wishes thank his (now retired) colleague and friend M. Vignati for his courtesy and the availability of refs. 1), 9).